\documentstyle[epsfig]{europhys}
\def\stars{\bigskip\centerline{***}\medskip}

\newif\ifboo \boofalse

\begin{document}
\euro{}{}{}{}
\Date{}
\shorttitle{ON THE DEPHASING TIME OF THE CHIRAL METAL}
\title{On the dephasing time of the chiral metal}
\author{J.J. Betouras}
\institute{
     Department of Physics, Theoretical Physics, 
     University of Oxford, \\
     1 Keble Road,
     Oxford, OX1 3NP,
     U.K.}
\rec{}{}
\pacs{
\Pacs{73}{23.-b}{Mesoscopic systems}
\Pacs{73}{20.Dx}{Electron states in low-dimensional structures 
(superlattices, quantum well structures and multilayers)}
\Pacs{73}{40.Hm}{Quantum Hall effect (integer and fractional)}
      }
\maketitle

\begin{abstract}
In the low-dimensional disordered systems the dephasing time and the 
inelastic scattering (out-scattering) time are in general different.
We show that in the case of the 
two-dimensional chiral metal which is formed at the
surface of a layered three dimensional system, which is 
exhibiting the integer
quantum Hall effect these two quantities are essentially the same and 
their temperature-dependence is $T^{-\frac{3}{2}}$. 
In particular we  show that the results obtained using the diagramatic
technique and the phase uncertainty approach introduced by A. Stern et 
al. (Phys. Rev. A {\bf{41}}, 3436 (1990)) for the out-scattering and
the dephasing time respectively, coincide. We furthermore
consider these quantities in the case of the three-dimensional 
chiral metal, where similar conclusions are reached.
\end{abstract}

Recently there has been much interest in the properties of the states
formed at the surface of a three dimensional (3D) structure which is
constructed by stacking together parallel two-dimensional systems which
exhibit the integer quantum Hall effect. The first study in this
direction,
by St\"ormer and collaborators \cite{stormer}  concluded that
strict two-dimensionality is not required for the observation of the
quantum Hall Effect and showed that the
conductivity in the direction of the field $\sigma_{zz}$
tends to zero in the bulk, as the temperature T approaches zero.

Subsequent
theoretical studies, first by Chalker and Dohmen \cite{john} and later 
by Balents and Fisher \cite{balents} were focused on 
the nature of the two dimensional states at the surface of the 
multilayered quantum Hall systems. In
particular, the former demonstrated the existence of the quantum Hall
regime in the layered 3D conductors in strong magnetic field as well as 
the surface states and predicted a metallic behaviour on the
$z$-axis. The latter, emphasized the absence of
any localization effects and mentioned that there are no
singularities in the density of states (DOS). The basic 
consequence of the absense of backscattering
due to the unidirectional electronic motion, is the Fermi liquid
behavior which is retained in the presence of interactions as opposed
to
the Luttinger liquid behavior in the interacting one-dimensional
electronic systems.

An experiment by Druist et al. \cite{druist} on
GaAs/Al$_{0.1}$Ga$_{0.9}$As multilayers allowed the observation of
electrons that are transported through the surface states, and indicated
the possibility of the 
suppression of  localization effects. In this experiment, the
conductance $g_{zz}$ approached a non-zero constant at very low
temperatures in the quantum Hall regime, and was
proportional to the circumference of the sample and not to
the area of the layers. This constituted a convincing evidence
that electrons are transported via the surface and not via the
bulk. 

The next phase of the experiments was to measure the
conductance fluctuations and to provide information on microscopic quantities
\cite{druist2}. In addition to that, 
conduction by surface states was considered
for the interpretation of experiments on bulk quantum Hall effect in
organic conductors \cite{hill1} and an inorganic quasi-two dimensional
conductor \cite{hill2}.

In a very recent theoretical study \cite{betouras} the
screened Coulomb  interaction
was taken into account in detail. Among other quantities, the
inelastic scattering rate (out- scattering rate in the mesoscopic
physics language) $\gamma = \hbar /\tau_{ee}$ was calculated using the
diagrammatic technique \cite{altshuler,fukuyama,blanter} 
and was found to vary as $T^{3/2}$. 
In general in the low dimensional interacting and disordered electron 
systems the out-scattering time is different from the dephasing time
$\tau_{\phi}$. 
The out-scattering time is the quantity that appears in the 
expression of the diffuson when the interaction effects are taken into
account and 
therefore it can be measured indirectly through the measurements
of the amplitude of the conductance fluctuations \cite{leestone,betouras}.
The dephasing time, on the other hand, is the time it takes for the
electrons to lose their phase coherence due to their interactions with
the environment and is the physical quantity in low-dimensional
disordered systems. 
The purpose of this letter is to show explicitly that in the system under
consideration the two times coincide and this is a special feature of the 
chirality of the system as we will discuss. In doing that we also
compare two different established calculational techniques. 

The Hamiltonian $H$ of the non-interacting system, acting on a wavefunction
$\psi_n(x)$ where $n$ labels the layer and $x$ is the chiral
direction,  is given by \cite{john,johnsondhi}:
\begin{eqnarray}
\nonumber
(H\psi)_{n}(x) =  -i v \hbar \partial_{x}
 \psi_{n}(x) 
-t[\psi_{n+1}(x)+\psi_{n-1}(x)] 
+ V_{n}(x) \psi_{n}(x)\,,
\label{schro} \\
\end{eqnarray}
where $v$ is the chiral velocity, 
$t$ is the interlayer tunneling energy, $a$ is the interlayer
spacing and $V_{n}(x)$ is a random potential arising from impurities and 
surface roughness. This random potential is chosen to be Gaussian
distributed with short-range correlations :
$\langle V_{n}(x)\rangle = 0$ and
$\langle V_{n}(x)V_{m}(x')\rangle = \Delta \delta_{nm} \delta(x-x')$.
The diffusion constant is $D=2(at)^2 v/\Delta$. 
Then the disorder averaged, retarded one-particle Green's function 
$G^R(\omega;{\bf k})$ in
real frequency and momentum space and also the 
diffuson at small momenta $K(\omega,{\bf k})$ are given by :
\begin{eqnarray}
G^R(\omega;{\bf k})&=& \frac{1}{
[\omega+i\Delta/(2\hbar v)-(\hbar v k_x-2t\cos(k_za)]} \\
K(\omega,{\bf k})&=& \frac{1}{(\hbar v)^{-1}[\hbar D {k_z}^2-i(\omega+\hbar v k_x)]}
\end{eqnarray}

\begin{figure}
\begin{center}
\epsfig{figure=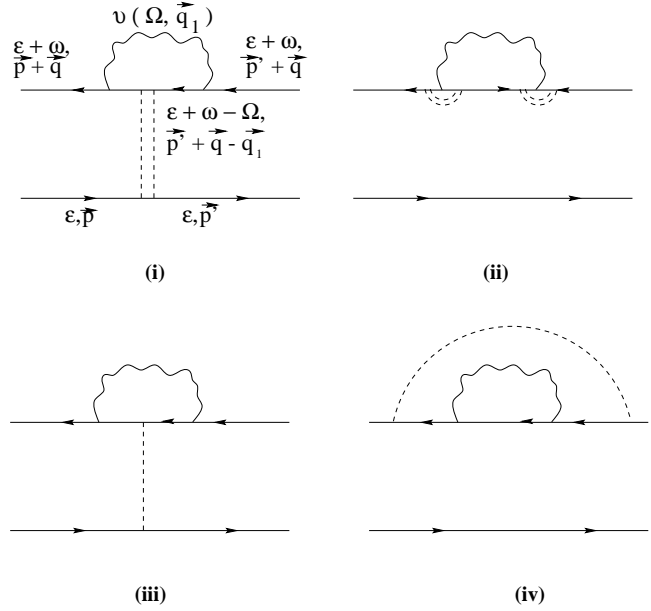,height=8cm}
\end{center}
\caption{The relevant diagrams for the calculation of the interaction 
block I. Straight lines correspond to electron propagators, wiggly lines
to Coulomb interactions, double dashed lines to diffusons and single
dashed lines to single impurity scattering.}
\label{fig:InScRate2_new}
\end{figure}

The screened Coulomb interaction, in Matsubara formalism and in Random
Phase Approximation (RPA), reads :
\begin{equation}
U_{eff}(i\Omega_n, {\bf q}) = U^{0}({\bf q})/ \epsilon =
U^{0}({\bf q})/ (1 + U^{0}({\bf q}) \Pi(i{\Omega_n}, {\bf q}) )
\end{equation}
where $\Omega_n = 2 \pi k_B T$ is the Matsubara frequency, 
$U^{0}({\bf q}) = 2 \pi e^2/q $ is the bare Coulomb interaction
in 2D, $\epsilon$ is the dielectric constant and $\Pi(i{\Omega}, {\bf
q})$ is the polarization which in both the RPA and the hydrodynamic approaches 
\cite{betouras} is : 
\begin{equation}
\Pi(i\Omega_n,{\bf q}) = n \frac{ -i \upsilon q_x {\rm sgn}(\Omega_n)+
D {q_z}^2}
{|\Omega_n| - i \upsilon q_{x} {\rm sgn}(\Omega_n) + D{q_z}^2}
\end{equation}
with $n=1/h \upsilon a$ being the density of states.     
In order to calculate $\gamma$  the interaction block technique can be
used \cite{blanter,fukuyama,betouras}:
\begin{equation}
\gamma = - \frac{1}{\pi n {\tau}^2} I
\end{equation}
where I is the interaction block. The relevant diagrams, up
to first order in the interaction, for this
particular problem, 
are those shown in Fig.~\ref{fig:InScRate2_new} together with their 
corresponding ones
with the interaction in the lower Green's function. The contribution 
comes from diagrams which do
not have interaction between different electronic lines. 

In order to do the summation over Matsubara frequencies, 
the contours of integration in the complex 
$z$-plane ($z = i \Omega$)
for the two first diagrams include a branch-cut at $z=0$ and all of
the four diagrams contain a ``boundary'' term at $z=i (\epsilon +
\omega)$ or $z=i\epsilon$. Combining all the terms and simplifying it 
under the assumption that the major contribution comes from small 
momenta, we arrive at the
expression for the out-scattering rate $\gamma$ :
\begin{eqnarray}
\nonumber
\gamma = \frac{1}{\pi \hbar n} 
\int \frac{d^2 q}{(2 \pi)^2}\int_{0}^{+\infty} dx [\coth(x/2k_BT) -
\tanh((x-\epsilon/)/2k_BT)] \times \\
\frac{1}{(x+\hbar \upsilon q_{x})^2 + (\hbar D {q_{z}}^2)^2} 
\frac{x (\hbar D {q_{z}}^2)^2}{\hbar^2{\upsilon}^2 {q_{x}}^2 + (\hbar
D q_{z}^2)^2} 
\end{eqnarray}
where the $\coth$ and $\tanh$ terms come from the Bose and the Fermi 
distribution function which appear in the $z=0$ branch-cut and 
the ``boundary'' terms respectively.
The main contribution comes from the small energy regime and
consequently the main term is the one with the $\coth$. 
We can rescale the momenta and x to :
\begin{equation}
\nonumber
x = k_BT X \;\;\;\;\;\;\;\;\;\;
\upsilon \hbar q_{x} = k_B T Q_{x} \;\;\;\;\;\;\;\;\;\;
\sqrt{\hbar D}q_{z} =  \sqrt{k_B T} Q_{z}
\end{equation}
The temperature dependence can be then extracted in the limit of small
temperatures and $\epsilon=0$, after performing the integration the full result is 
\cite{betouras}:
\begin{equation}
\gamma = 1.5 \frac{a}{D^{1/2}} 
\left(\frac{k_{\rm B}T}{\hbar}\right)^{3/2}\
\end{equation}
This $T$-dependence can be understood quantitatively in relation to the 
general argument that $\gamma
\propto 1/n_d {L_T}^{d}$ where $d$ is the dimensionality of the
system and $L_T$ is the characteristic thermal length 
which acts as a cut-off. In the case of the chiral metal the
characteristic 
thermal lengths are different in the two dimensions due to
the anistropic behavior. In that case
${L_T}^{d} = L_{T,x}L_{T,z}$, and, since $ L_{T,x} \propto (\upsilon /T)$ 
and $L_{T,z} \propto \sqrt{D/T}$ then the rate $\gamma \propto T^{3/2}$.

We now follow the general procedure for the computation
of $\tau_{\phi}$ which was introduced in the pioneering work 
\cite{stern} by Stern, Aharonov and Imry (from now on in this paper 
referred as SAI). 
The basic physical picture which was put forward and analysed in this work 
(supplementing the important work of \cite{altshuler3}) 
was the thought interference experiment
of an electron which can choose a left or right path before 
interfering with itself. The description of the electron with 
the wave function $\psi({\bf r}) = \psi_{L}({\bf r}) \otimes
\psi_{R}({\bf r})$ allows the 
independent interaction of each part (Left/Right) of the wave function 
with its environment. This leads naturally to dephasing and the key
observation is that the phase uncertainty  accumulated by the electron
is given by twice the probability $P(t)$ that the environment alters its state
due to the interaction with the electron. This allows for a
quantitative treatment of the dephasing time; from the phase
uncertainty we can get the information on the temperature dependence
of the $\tau_{\phi}$  which is now defined as the time where
$P(\tau_{\phi}) \simeq 1$.  
It is evident that this procedure does not  allow for the accurate 
calculation of the prefactors.

If we consider up to second order terms 
in the interaction, then the probability is given by :
\begin{eqnarray}
\nonumber
P(t_0) = \sum_{|\alpha\rangle \neq |0\rangle} \int_0^{t_0} dt
\int_0^{t_0} dt' \langle 0| V({\bf r_1}(t),t) - V({\bf r_2}(t),t) |\alpha
\rangle \\ 
\langle \alpha | V({\bf r_1}(t'),t') - V({\bf r_2}(t'),t') |0 \rangle 
\end{eqnarray}
where $V({\bf r},t)$ is the Coulomb interaction between the observed electron
and the rest of the electrons which consist the 
environment, and ${\bf r}_1(t)$, ${\bf r}_2(t)$ are the two different 
paths (of equal length) under consideration. 

Then, using the fluctuation-dissipation
theorem, $P$ is given by \cite{imry}:
\begin{equation}
P(t_0) = \frac{1}{\hbar}\int_{0}^{t_0} dt \int_{0}^{t_0} dt'
\int_{-\infty}^{\infty} d\omega {\rm coth}(\frac{\omega}{2 k_B T}) \int
\frac{d^2 q}{( 2 \pi)^2} \frac{2 \pi e^2}{q} {\rm Im}(
\frac{1}{\epsilon}) \exp[i \omega (t- t') + i {\bf q} \cdot
[{\bf r}_1(t) - {\bf r}_2(t')] ] 
\end{equation}
The above expression contains only one of the four possible terms 
which arise when we take into account the phase fluctuations of the 
two paths ( $< (\delta \phi_1 - \delta \phi_2)^2 > =  < (\delta
\phi_1)^2> + 
<(\delta \phi_2)^2> - 2 <\delta \phi_1 \delta \phi_2 > $ ).
The other terms contain the phases :
\begin{equation}
\nonumber
\exp( i {\bf q} \cdot [{\bf r}_1(t) - {\bf r}_1(t')])
\;\;\;\;\;\;\;\;\;\;
\exp(i {\bf q} \cdot
[{\bf r}_2(t) - {\bf r}_2(t')]) \;\;\;\;\;\;\;\;\;\;
\exp(i {\bf q} \cdot
[{\bf r}_2(t) - {\bf r}_1(t')] 
\end{equation}
For the chiral metal :
\begin{equation}
\nonumber
{\bf q}\cdot [{\bf r}_1(t) - {\bf r}_2(t')] 
= q_x \upsilon (t-t') + q_z ( z_1(t) - z_2 (t'))
\end{equation}
${\rm Im}(1/\epsilon)$ is calculated using Eq.(4) and (5) 
and the factor $\coth$ restricts the
integration to small frequencies ($-k_B T< \omega < k_B T$) and can be
expanded. 
Since the only $\omega$ dependence is retained in the exponential
$\exp(i\omega (t-t'))$
in the limit of small $q$ (in fact the only mathematical
assumption in the calculation is that $1 + 2\pi n e^2/q \simeq 2\pi n
e^2/q $), and under the physical
assumption that the duration of the experiment is longer than $1/k_B
T$, the integration over frequency can be well approximated by
$\delta(t - t')$. The integration over $t'$ then is trivial and the 
integration over $q_x$ can also be performed easily by a residue
integration (note that the terms that come from the bare Coulomb
interaction and contain $q$ drop out of the calculation under the
mathematical assumption made above).
If at this stage we take into account all four terms then the
expression that remains to be evaluated is :
\begin{equation}
P(t_0) = \frac{1}{n\hbar}  
\int_0^{t_0} dt k_B T \int dq_z {\rm sin}^2[\frac{q_z}{2} (z_1(t) - z_2(t))]
\end{equation}
The integrand of this expresion is identical to the integrand obtained
in the case of the 3D, disordered, ordinary metal in
spherical coordinates
after performing the angular integrations \cite{imry} as we will
comment below.
The upper cut-off in the integration over $q_z$ is then 
$q_{z,max} = \sqrt{k_B T/D} = {L_{T,z}}^{-1}$. In addition to that, and due to
the diffusive nature of the motion in the $z$-direction, we can approximate 
$|z_1(t) - z_2(t)| \sim \sqrt{Dt}$.
Taking into account all the ingredients the final result is :
\begin{equation}
\nonumber
P(t_0) \simeq  t_0 \frac{a}{\sqrt{D}} (k_B T)^{3/2}
\end{equation} 
Since $P(t_0) \simeq O(1)$ then  
$\tau_{\phi} \propto (k_B T)^{-3/2}$, i.e. it exhibits the same $T$-
dependence as $\tau_{ee}$. 

We recall that in the ordinary (non-chiral) 2D dirty metals the 
out-scattering time $\tau_{ee}^{-1} \sim T |{\rm log}T|$ \cite{altshuler}. 
The reason for the logarithmic correction is that the dominant contribution 
to the scattering comes from 
processes with small energy transfers. On the other hand these,
processes are not the dominant ones in the calculation of the
dephasing time and this fact leads to the difference in the temperature
dependence of the two quantities. In the case of the chiral metal though the
small energy transfer processes are not the dominant ones either for the
dephasing or for the out-scattering time and therefore these quantities
coincide. 

If we consider the equivalent calculations for a 3D chiral
metal (conceived as coupled chiral chains in two directions), where a 
ballistic motion occurs in one direction and a
diffusive motion in the other two, then the polarization
is  modified to :
\begin{equation}
\Pi(i\Omega_n,\vec{q}) = n \frac{-i\upsilon q_x {\rm sgn(\Omega_n)} + D
{q_y}^2 + D' {q_z}^2 }{|\Omega_n| -i\upsilon q_x {\rm sgn(\Omega_n)} + D
{q_y}^2 + D' {q_z}^2 }
\end{equation} 
where we assume different diffusion constants for the two non-chiral
directions. Taking now into account that $U^{0}(\vec{q})= 4\pi
e^2/q^2$, $\gamma$ becomes :
\begin{equation}
\gamma = c   \frac{ab}{\sqrt{D D'}}(\frac{k_B T}{\hbar})^2 
\end{equation} 
with $a$ and $b$ being the interchain distances in the two perpendicular
directions to the chains and $c$ a constant of $O(1)$. 
The SAI approach then yields the same temperature dependence for the 
dephasing time for the same reasons as those described above.
Note the accidental
similarity in the temperature dependence of the dephasing rate  
of the 2D chiral metal with the
3D ordinary disordered metal as well as the similarity in the
temperature dependence of the inelastic scattering rate of the 
3D chiral metal with that of a 3D clean metal.

The conclusion of the presented work is that the novel electronic
system (chiral metal) offers a chance to test the different approaches
of calculating the characteristic times of a mesoscopic system. 
We considered the system in the regime  where the inter-edge
tunneling time $\tau_{\perp}$ satisfies the condition
$\tau_{\perp} \ll \tau_{\phi}$.
The dephasing time, 
although it has a transparent physical meaning, is
difficult to be measured in the chiral metal but
the temperature dependence of the out-scattering time can be extracted
through the amplitude of the conductance fluctuations.
We demonstrated that the two times have the same temperature
dependence  as a result of the
subdominence of the small energy transfer processes. The 
general quantitative arguments given in \cite{altshuler} which can 
be formally represented by the interaction block technique are confirmed
in  this case  and give identical results with the
procedure introduced by SAI. The interplay of electron-electron
interactions and disorder in this low-dimensional system does not lead
to the same consequences as in the ordinary dirty metallic systems due
to the absence of self-intersecting paths in the electronic motion.

\stars

The author is grateful to John Chalker for 
helpful discussions and careful reading of the manuscript. This work has been
supported in part by the European Union under TMR FMBICT972304 and in
part by EPSERC under grant No GR/J8327.

\end{document}